\documentclass[twocolumn,showpacs,10pt,prb]{revtex4}
\usepackage{amssymb}
\usepackage{graphics}
\usepackage{epsfig}

\begin{document}

\title{Composition and field tuned magnetism and superconductivity in Nd$%
_{1-x}$Ce$_{x}$CoIn$_{5}$}
\author{Rongwei Hu,$^{1,2}$ Y. Lee,$^{3}$ J. Hudis,$^{4}$ V. F. Mitrovi{c,}$%
^{2}$ and C. Petrovic$^{1}$}
\affiliation{$^{1}$Condensed Matter Physics, Brookhaven National Laboratory, Upton New
York 11973-5000 USA\\
$^{2}$Physics Department, Brown University, Providence RI 02912\\
$^{3}$Department of Earth System Sciences, Yonsei University, Seoul 120749,
Korea\\
$^{4}$Department of Physics and Astronomy, Johns Hopkins University,
Baltimore, MD 21218}
\date{\today}

\begin{abstract}
The Nd$_{1-x}$Ce$_{x}$CoIn$_{5}$ alloys evolve from local moment magnetism
(x = 0) to heavy fermion superconductivity (x =1). Magnetic order is
observed over a broad range of $x$. For a substantial range of $x$ (0.83 $%
\leq $ x $\leq $ 0.95) in the temperature - composition phase diagram we
find that superconductivity may coexist with spin - density wave magnetic
order at the Fermi surface. We show that a delicate balance betwen
superconducting and magnetic instabilities can be reversibly tuned by both
the Ce/Nd ratio and magnetic field, offering a new and unique model
electronic system.
\end{abstract}

\maketitle

\section{Introduction}

In contrast to simple metals such as Pb, phase diagrams of unconventional
superconductors often show a multitude of electronic phases of matter. In
particular, magnetic order is ubiquitous in cuprate oxides and heavy fermion
superconductor (HFSC) phase diagrams alike, as well as in some ruthenates
and cobaltates. The proximity, competition, or coexistence of two distinct
types of electronic order at the Fermi surface raised speculations that
their driving mechanisms could be closely related.\cite{Fisk1}$^{,}$ \cite%
{Demler} Exotic superconductivity in heavy fermion materials \cite{Fisk2}
usually appears near the quantum critical point \cite{Senthil} where the
magnetic ordering temperature is tuned to T =0 by a variety of external
parameters. For example annealing, composition or magnetic field.\cite%
{Trovarelli} However, the most frequently used tuning parameter in HFSC is
pressure.\cite{Yuan} This is in variance with the cuprate family where the
interplay of superconductivity and magnetism is balanced by adjusting the in
plane charge density of the CuO$_{2}$ layers. The discovery of pressure -
induced superconductivity in CeIn$_{3}$\cite{Mathur} stimulated exploratory
synthesis of AuCu$_{3}$ superstructures by Fisk, where the magnetic entropy
might be further suppressed by crystallographic arguments.\cite{Fisk3}
Indeed, superconductivity in CeRhIn$_{5}$,\cite{Rh} and CeIrIn$_{5}$,\cite%
{Ir} was soon discovered, as well as coexistence of heavy fermion
superconductivity and magnetism in CeRh$_{1-x}$Ir$_{x}$In$_{5}$.\cite%
{Pagliuso}$^{,}$\cite{Christianson} CeCoIn$_{5}$,\cite{Co} the \textit{primo}
compound in the 115 family of HFSC, has also been recently hole doped to an
antiferromagnetic ground state by Cd substitution on the In (1) site.\cite%
{Pham} Despite a few examples of structurally tuned superconductivity in
HFSC, the lack of a predictive theory suggests that there is still no clear
understanding of how the delicate interplay of various degrees of freedom in
these materials stabilizes superconducting, or a magnetic ground state.

The large quasiparticle mass enhancement in CeCoIn$_{5}$ is reflected in two
large nearly cylindrical pieces of the Fermi surface and smaller 3D hole
pockets.\cite{Settai} Due to the pronounced $\overrightarrow{k}$ - space
inhomogeneity, hybridization of the $3d$ electrons of Co with the $5p$
electrons of In results in a small density of states at the Fermi energy,
implying partially quasi - 2D electronic structure.\cite{Settai2} By
exploring a well defined non - hybridizing local moment such as Nd, we were
able to continuously tune the coupling in the lattice and consequently
quasiparticle mass enhancement between 4f ions and conduction electrons. As
a result, we have obtained a rich phase diagram. The magnetic ground state
in Nd$_{1-x}$Ce$_{x}$CoIn$_{5}$ smoothly evolves from local moment magnetism
(LMM) on the Nd - rich side to HFSC on the Ce - rich side. Superconductivity
coexists with other forms of electronic order, most likely magnetic in
nature, in samples for Nd concentrations between $x=0.78$ and $x=0.98$. We
demonstrate that the delicate balance between coexisting ordered states can
be smoothly tuned by magnetic fields near the magnetic - superconducting
boundary.

\section{Methods}

Single crystals of Nd$_{1-x}$CexCoIn$_{5}$ were grown from an excess In
flux. Magnetic susceptibility, specific heat and resistivity measurements
were performed in a Quantum Design MPMS XL 5 and PPMS - 9 instruments
respectively. Single crystals were thoroughly ground to a fine powder for
structural measurements. High resolution synchrotron powder X - ray patterns
were taken at beamline X7A of National Synchrotron Light Source at
Brookhaven National Laboratory. Monochromatic synchrotron x-ray and
gas-proportional position-sensitive detector were used to measure the powder
diffraction data. Rietveld refinements were performed using GSAS.\cite{GSAS}
The samples were manually aligned to measure the magnetization, heat
capacity, or resistivity for fields applied along the appropriate axis. M/H
polycrystalline averages were calculated as $\chi (T)=[2\chi _{ab}(T)+\chi
_{c}(T)/3]$ and were used to obtain the high temperature effective moments.
Temperature dependent magnetic susceptibility was used to estimate the
relative ratio of Ce/Nd ions. At high temperatures ranging from 150 K to 350
K, H/M curves were fit with Curie - Weiss (CW) law, $\chi (T)=C/(T-\Theta ), 
$ where C is the Curie constant and $\Theta $ is the Weiss temperature.
Electrical contacts for in - plane resistivity measurements were made with
Epotek-H20E silver epoxy on thin plate-like crystals whose dimensions were
measured by optical microscope with 10$\mu $m resolution. The samples were
previously etched in diluted HCl for several hours and thoroughly rinsed in
ethanol in order to remove excess In.

\section{Structural Characterization and Phase Purity}

Since both NdCoIn$_{5}$ and CeCoIn$_{5}$ are grown using identical
temperature profiles and an identical ratio of starting materials Nd (Ce),
Co and In,\cite{Co}$^{,}$\cite{Jacob} one would expect a smooth change of
the lattice parameters in the Nd$_{1-x}$Ce$_{x}$CoIn$_{5}$ alloy series.

\begin{figure}[t]
\centerline{\includegraphics[height=2.2in]{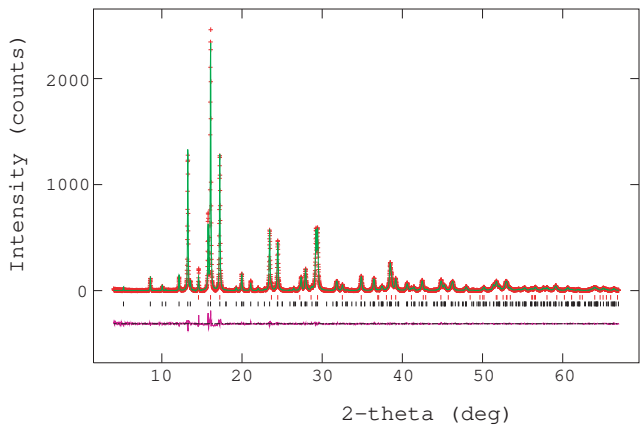}} 
\vspace*{-0.2cm}
\caption{Synchrotron powder X - ray diffraction data of Nd$_{0.2}$Ce$_{0.8}$%
CoIn$_{5}$. For all x, only the HoCoGa$_{5}$ structure was detected, in
addition of several small peaks of pure In from the flux.}
\end{figure}

Indeed, this is confirmed by high resolution structural measurements which
showed that the samples crystallized in tetragonal HoCoGa$_{5}$ structure
without any additional peaks introduced by Ce alloying (Fig. 1). Selected
regions of the powder X - ray spectra showed monotonic evolution and uniform
sharpness independent of $x$ for both [111] and [003] peaks. This implies
that Ce uniformly substitutes Nd with the increase of x and that our samples
are indeed alloys rather than a mixture of intergrown compounds NdCoIn$_{5}$
and CeCoIn$_{5}$. The lattice parameters increase smoothly with Ce
substitution in accordance with Vegard's law (Fig. 2b).

\begin{figure}[t]
\centerline{\includegraphics[height=4in]{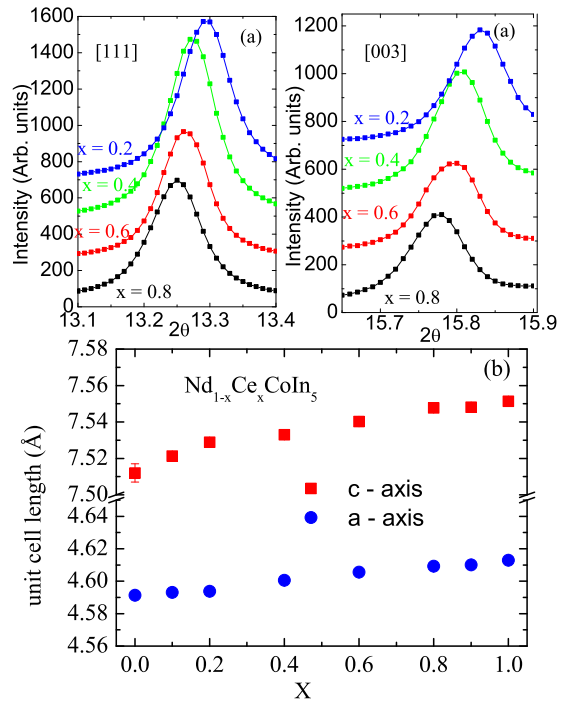}} 
\vspace*{-0.2cm}
\caption{Selected regions of the synchrotron powder X - ray diffraction data
for Nd$_{1-x}$Ce$_{x}$CoIn$_{5}$. Both [111] and [003] peaks shift uniformly
with Ce substitution. (b) Tetragonal lattice parameters $\widehat{\mathit{a}}
$ (red symbols) and $\widehat{\mathit{c}}$ (blue symbols)}
\end{figure}

Magnetic susceptibility measurements did not detect magnetically ordered
second phases, such as NdIn$_{3}$ or CeIn$_{3}$ (Fig. 3). The cubic
compounds CeIn$_{3}$ and NdIn$_{3}$ order antiferromagnetically at 10 K and
6.3 K respectively.\cite{Mathur}$^{,}$\cite{Amara} Magnetic ordering in
NdCoIn$_{5}$ is depressed smoothly with the increase of Ce, and we were able
to follow the characteristic peak in M/H (signature of the onset of the
magnetic order) down to lowest temperature of our magnetic measurement, T =
1.8 K for x = 0.6. Curie - Weiss analysis of polycrystalline magnetic
susceptibility average at high temperatures showed at most a 4\% deviation
from the nominal ratio of Ce$^{3+}$ and Nd$^{3+}$ moments (Table 1). As
expected, the highest uncertainty in the nominal concentration of Ce (x) is
in the middle of the alloy series, for the highest chemical disorder.
Combined together, these results demonstrate that Ce uniformly substitutes
Nd in the entire doping range of Nd$_{1-x}$Ce$_{x}$CoIn$_{5}$, with a
maximum $\Delta x$ = 0.04.

\begin{table*}[tbp]
\caption{High temperature magnetic moment in Nd$_{1-x}$Ce$_{x}$CoIn$_{5}$
alloy series}%
\begin{tabular}{|l|l|l|l|}
\hline\hline
x & Measured $\mu _{eff}$($\mu _{B}$) & Expected$\mu _{eff}$($\mu _{B}$) & 
Error (\%) \\ \hline\hline
1 & 2.59(1) & 2.54 & 2 \\ \hline\hline
0.85 & 2.67(6) & 2.69 & 0.8 \\ \hline\hline
0.83 & 2.73(1) & 2.71 & 0.8 \\ \hline\hline
0.8 & 2.72(1) & 2.75 & 1 \\ \hline\hline
0.6 & 2.83(1) & 2.96 & 4 \\ \hline\hline
0.5 & 2.96(1) & 3.07 & 3 \\ \hline\hline
0.2 & 3.37(2) & 3.40 & 0.8 \\ \hline\hline
1 & 3.7(1) & 3.62 & 2 \\ \hline
\end{tabular}%
\end{table*}

\section{Results}

M/H data of Nd$_{1-x}$Ce$_{x}$CoIn$_{5}$ alloy series show substantial
anisotropy (Fig. 3) at low temperatures and signature of magnetic order at T$%
_{N}$ = 9 K. For field applied along the $\widehat{a}$-axis, $H||\widehat{a}%
, $ the magnetic transition is rather broad. On the other hand for $H||%
\widehat{c},$ the transition is relatively sharp implying complex magnetic
order.

\begin{figure}[t]
\centerline{\includegraphics[height=4in]{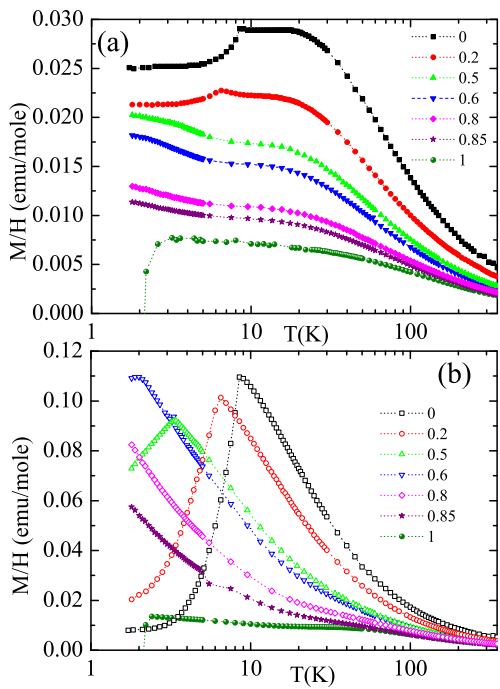}} 
\vspace*{-0.2cm}
\caption{Magnetic properties of Nd$_{1-x}$Ce$_{x}$CoIn$_{5}$ alloy series in
1kOe for a)H $\uparrow \uparrow $ $\widehat{a}$ and b)H $\uparrow \uparrow $ 
$\widehat{c}$ axis. Magnetic susceptibility shows decrease with Ce
substitution, with easy axis along the crystalline $\widehat{c}$ axis. The
characteristic signature of antiferromagnetic order is observed down to $%
x=0.6$ above $T=1.8K$ for $\protect\chi _{C}$(T). The smooth evolution of $%
\protect\chi _{C}$(T) with x could indicate that magnetic fluctuation
spectrum along the $\widehat{c}$ axis is more relevant for tuning of the
ground state. }
\end{figure}

The thermodynamic properties of Nd$_{1-x}$Ce$_{x}$CoIn$_{5}$ are shown in
Fig. 4. The magnetic ordering temperature in NdCoIn$_{5}$ is smoothly
depressed from T$_{N}$ = 9 K with increased Ce concentration to 2.0 K by $%
x=0.6$ where (C-C$_{latt}$)/T shows a broad peak (Fig. 4a). Interestingly,
the magnetic ordering transition becomes sharper as $x$ is further tuned
towards the superconducting boundary for $x=0.7$. At $x\geq 0.78$
superconductivity emerges. With increasing $x$, the \.{T}$_{C}$ increases to
the bulk T$_{C}$ = 2.3 K of CeCoIn$_{5}$. The magnetic entropy released upon
emergence of LMM (Fig. 4a inset) scales with Nd concentration up to $x=0.4$,
implying that Ce ions do not play a direct part in the formation of the LMM
ground state. In the HFSC state for $0.78\leq x\leq 1$, the magnetic entropy
at 5K is essentially invariant to changes of Ce/Nd ratio. However, a
hallmark of heavy fermion magnetism is observed for $x\geq 0.5$. This can be
seen on the example of $x=0.6$ where only approximately $\sim 0.2$ Rln2 is
released below the magnetic ordering transition at 2 K. In the
superconducting region at the Ce -rich side (Fig. 4b) additional
thermodynamic anomalies $A$, $B$ and $C$ emerge in (C-C$_{latt}$)/T below
the superconducting transition T$_{C}$ for $x=0.95$, $x=0.9$, but above T$%
_{C}$ for $x=0.83$.

\begin{figure}[t]
\centerline{\includegraphics[height=4in]{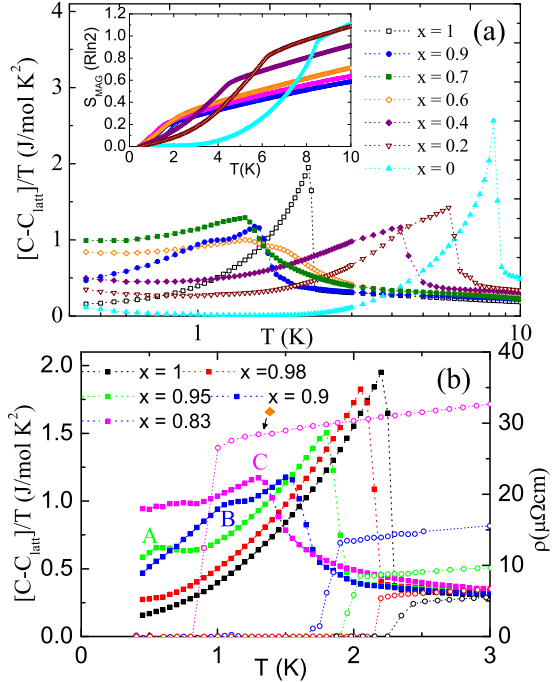}} 
\vspace*{-0.2cm}
\caption{(a): Thermodynamic properties of Nd$_{1-x}$Ce$_{x}$CoIn$_{5}$.
Specific heat of LaCoIn$_{5}$ was taken as the estimate of the lattice
specific heat C$_{latt}$. Magnetic entropy (inset) obtained from integral
(C-C$_{latt}$)/T in the same temperature range. (b): Thermodynamic and
transport properties in the superconducting state, measured on the same
sample for each Ce concentration $x$ minus the lattice. The lattice
resistivity was estimated by taking values of electrical resistivity of
LaCoIn$_{5\text{.}}$ With decrease of the Ce/Nd ratio, the superconducting
temperature T$_{C}$ is suppressed. The additional thermodynamic anomaly
increases in temperature (A $\longrightarrow $ B $\longrightarrow $ C), as
seen in the heat capacity data (full symbols). In - plane resistivity data
(open symbols) are shown to identify superconducting transition.}
\end{figure}

By examining electronic transport we obtain further evidence for the
presence of distinct types of electronic matter in the phase diagram.
Temperature dependent electrical resistivities normalized to their value at
300K (Fig. 5 a,b) show rather small loss of the spin disorder scattering
below Neel temperature in NdCoIn$_{5}$. This is consistent with the LMM -
type of order in rare earth intermetallic compounds where local moments do
not become part of Fermi surface upon cooling. By $x=0.5$ the electronic
scattering strongly increases due to a single ion Kondo - type interaction
and a logarithmic contribution of hybridizing Ce$^{3+}$ ions submerged in
the Fermi sea. Eventually for $x\geq 0.5$ a broad coherence peak develops in
the lattice. This marks the emergence of an additional Kondo energy scale
arising from the coherence in the heavy fermion lattice. As expected, the
coherence temperature T$_{COH}$ increases with the increase of Ce$^{3+}$
ions. Looking from the Ce - rich side, the onset of the superconducting
transition in CeCoIn$_{5}$ is depressed to 0.9 K and magnetic scattering
increases by $x=0.83$ (Fig. 4b).

\begin{figure}[t]
\centerline{\includegraphics[height=4in]{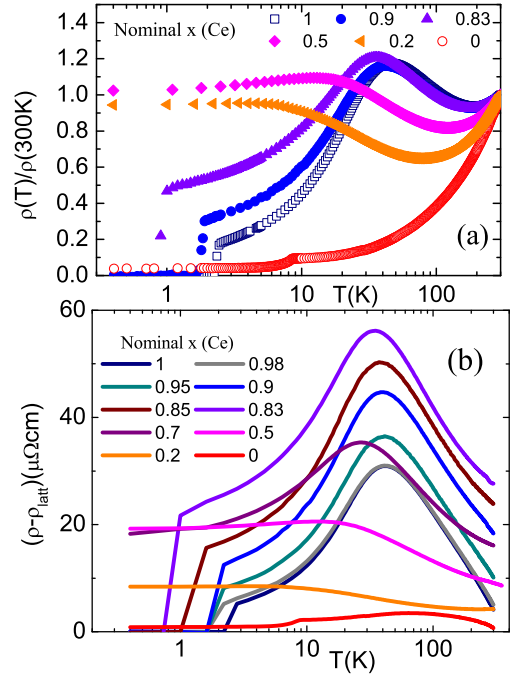}} 
\vspace*{-0.2cm}
\caption{Results of resistivity measurements on Nd$_{1-x}$Ce$_{x}$CoIn$_{5}$
with current parallel to the [100]. (a): Resistivity curves normalized to
their value at 300 K. (b): Magnetic contribution to the electrical
resistivity ($\protect\rho $ - $\protect\rho _{lattice}$). The lattice
resistivity was estimated by taking values of electrical resistivity for
LaCoIn$_{5}$.}
\end{figure}

\section{Discussion}

Heat capacity in the HFSC state for $x\leq 0.95$ reveals two thermodynamic
anomalies. Only one (C-C$_{latt}$)/T discontinuity, though, corresponds to
the superconducting transition (Fig. 4b). One possible explanation for this
would involve sample inhomogeneity and a distribution of the doping
concentration. This however is very unlikely due to the low residual
resistivity, the clean high resolution synchrotron powder X - ray
diffraction pattern with uniform sharpness of peaks for the whole range of $%
x $ and the compositional dependence of lattice parameters $\widehat{a}$ and 
$\widehat{b}$ in accordance with Vegard's law (Fig.1). Furthermore, one
would expect the highest degree of metallurgical disorder around the middle
of the doping range, for $x=0.5$, and not near the Ce end. Finally, in
superconducting materials with \textit{metallic} type of bonding,
metallurgical inhomogeneity would shift the bulk T$_{C}$ of the \textit{whole%
} sample, not only a fraction of the sample. For example, in CeCu$_{2.2}$Si$%
_{2}$ this is indeed seen in high resolution studies of the local structure.%
\cite{Louca} Moreover, as seen in the heat capacity data for $x=0.1$, the
two thermodynamic anomalies are comparable in size. Thus secondary phases
would have been easily detected in the analysis of powder X - ray spectra.
Furthermore, as will be evident in Fig. 7, the smooth evolution of $T_{C}(x)$
and $T_{M}(x)$ argues against real space inhomogeneity. Our results imply
that magnetic and superconducting phases coexist in the phase diagram,
similarly to the situation in Cd - doping of CeCoIn$_{5}$.\cite{Pham}

\begin{figure}[t]
\centerline{\includegraphics[height=4in]{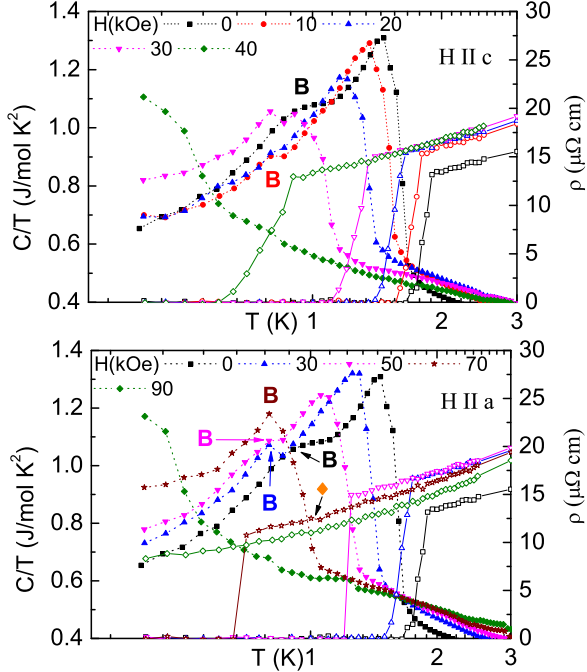}} 
\vspace*{-0.2cm}
\caption{Thermodynamic and transport properties near magnetic -
superconducting boundary, $x=0.1$, for H $\uparrow \uparrow $ $\widehat{c}$
- axis (a) and H $\uparrow \uparrow $ $\widehat{a}$ - axis (b). For each $x$%
, C/T and $\protect\rho $ data were taken on the same sample. The crossover
from positive to negative magnetoresistance for H $\uparrow \uparrow $ $%
\widehat{a}$ - axis suggests different nature of critical magnetic
fluctuations below H$_{C}$ = 50 kOe and above H$_{C}$. Superconducting
transition T$_{C}$ corresponds to thermodynamic anomaly at higher
temperature (below H$_{C}$) and at lower temperature (above H$_{C}$) for H $%
\uparrow \uparrow $ $\widehat{a}$ - axis. A small feature in resistivity
around 1 K for H = 70 kOe could indicate opening of the partial gap at the
Fermi surface in higher fields.}
\end{figure}

An alternative explanation involves two superconducting energy scales on
different parts of the Fermi surface and negligible interband scattering.
This could explain low temperature anomalies $A,B$ for $x=0.95$ and $x=0.9$
since zero resistivity is achieved at the high temperature transition and is
maintained through the low temperature transition. However, this scenario
cannot explain thermodynamic anomaly $C$ above the superconducting
transition in \ Nd$_{0.17}$Ce$_{0.83}$CoIn$_{5}$ (Fig. 4b). We speculate
that low temperature discontinuities $A$ and $B$, high temperature
discontinuity $C$ as well as sharp transition in [C-C$_{latt}$]/T (Fig.
4(a)) for x = 0.7 may be connected with some form of Fermi surface
instability appearing concurrently on a different part of the Fermi surface.
In what follows we show that this instability as well as the boundary
between superconductivity and magnetism can be tuned by a magnetic field, in
addition to composition $x$. Application of magnetic field reverses
remarkable behavior seen with progressive decrease in the Ce/Nd ratio,
tuning the boundary between the two ordered states.

\begin{figure}[t]
\centerline{\includegraphics[height=2.8in]{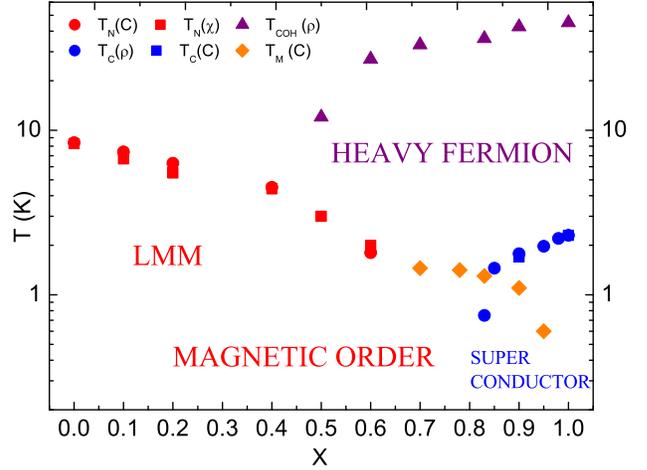}} 
\vspace*{-0.2cm}
\caption{Phase diagram of Nd$_{1-x}$Ce$_{x}$CoIn$_{5}$ at H = 0:
superconducting T$_{C}$ (blue symbols), antiferromagnetic Neel temperature
(red symbols), coherence temperature of the Kondo lattice (purple symbols).
Orange symbols represent second broad thermodynamic anomaly in C/T (below T$%
_{C}$) where resistivity shows small upturn (above T$_{C}$ and for x = 0.7
at H = 0). The local moment antiferromagnetism of NdCoIn$_{5}$ smoothly
changes to heavy fermion antiferromagnetic state around $x=0.5$. Magnetism
and superconductivity meet around $x=0.8$. The Neel ordering temperature
monotonically decreases for $0\leq x\leq 0.6$. For $x\geq 0.7$ heat capacity
anomaly most likely corresponds to spin density wave type of magnetic order,
deep in the superconducting state.}
\end{figure}

The magnetic field depresses both $T_{C}$ and $B$ in Nd$_{0.9}$Ce$_{0.1}$CoIn%
$_{5}$ (Fig 6). The suppression is rather anisotropic. For a field applied
along the $\widehat{c}$- axis superconducting anomaly in heat capacity and
increasingly broad resistivity transition are suppressed to 0.5 K in 40 kOe.
The anomaly $B$ is observed only below 10 kOe, merging into a single
thermodynamic transition for higher fields above 0.4 K, the lowest
temperature of our measurement. On the other hand, for a field applied along 
$\widehat{a}$- axis both thermodynamic anomalies show quite different field
dependence. Superconducting T$_{C}$ at 30 kOe - defined by the simultaneous
onset of zero resistivity and the start of the heat capacity anomaly - is
suppressed to 1.4 K. The anomaly $B$ is observed below the T$_{C}$, just as
we observe in H = 50 kOe. In contrast to H = 0 and H = 30 kOe, zero
resistivity in H = 50 kOe corresponds to midpoint, rather than the onset of
the heat capacity anomaly. In H = 70 kOe, Nd$_{0.1}$Ce$_{0.9}$CoIn$_{5}$
electronic matter becomes equivalent to the $x=0.83$ sample at H = 0 kOe:\
the main heat capacity transition is now above superconductivity. Electronic
scattering for field applied in tetragonal plane (Fig. 3b) first increases
and then decreases in the normal state as the magnetic field is tuned
through H$_{C}$ = 50 kOe. The resistivity transition in H $\geq $ H$_{C}$ is
sharper implying a magnetic field induced phase transition for H $\uparrow
\uparrow $ $\widehat{a}$ -axis. The size of the superconducting anomaly in
(C-C$_{latt}$) decreases in field relative to the size of additional
thermodynamic anomaly B below T$_{C}$ at H = 0 for field applied along both
crystalline axes. There is no loss of spin disorder scattering at the main
heat capacity transition T$_{M}$ for both x = 0.1 sample in H = 70 kOe (Fig.
6b) and for x = 0.83 sample in H = 0 (Fig. 4b). Instead, a very small upturn
is observed in Fig. 4(b) (x = 0.83, H = 0) and in Fig. 6(b) (x = 0.1, H = 70
kOe) in resistivity (denoted by orange diamonds), a signature of the partial
gapping of the Fermi surface. This is reminiscent of an itinerant
spin-density-wave type - transition based on Fermi-surface nesting in a
heavy-electron band observed in Ce(Ru$_{0.85}$Rh$_{0.15}$)$_{2}$Si$_{2}$.%
\cite{Murayama}

It appears that Nd$_{1-x}$Ce$_{x}$CoIn$_{5}$ can support simultaneously
multiple electronic ordering states. Coexistence of magnetism in
superconductivity in Ce 115 superconductors has been induced by pressure in
CeRhIn$_{5}$ or by composition in CeRh$_{1-x}$Ir$_{x}$In$_{5}$, CeCo(In$%
_{1-x}$Cd$_{x}$)$_{5}$ and CeRh$_{1-x}$Co$_{x}$In$_{5}$.\cite{Rh}$^{,}$\cite%
{Pagliuso}$^{,}$\cite{Pham}$^{,}$\cite{Zapf} \ The low temperature specific
heat and resistivity data taken in Nd$_{0.1}$Ce$_{0.9}$CoIn$_{5}$ at H = 0
and in Nd$_{0.17}$Ce$_{0.83}$CoIn$_{5}$ at H = 70 kOe imply that T$_{C}$ and
T$_{M}$ involve instabilities at different parts of the Fermi surface. The
balance between these states is tuned only by degree of hybridization in the
Kondo Lattice (Ce/Nd ratio) and by magnetic field. Though the presence of
additional magnetic transition made it difficult to estimate accurate values
for the electronic heat capacity coefficient $\gamma $, by taking $\gamma $
= [C-C$_{latt}$(T)]/T above T$_{C}$, we observe that the large jump at
ambient pressure in the specific heat of CeCoIn$_{5}$ $\Delta $C/C(T$_{C}$)
= 4.35 is reduced as more Nd enters into the matrix. For x = 0.98 we observe 
$\Delta $C/C(T$_{C}$) = 3.79 and by x = 0.1 this ratio decreases to 1.76.
Nevertheless, our results may indicate general trend that increased Nd
concentration decreases electron - boson coupling strength.\cite{Bang}

In the standard paradigm, the suppression of heavy fermion antiferromagnetic
order leads to superconductivity around the quantum critical point.\cite%
{Millis}$^{,}$\cite{Steglich2} One aspect of the reduced Ce/Nd ratio in the
lattice is a negative pressure. Using the bulk modulus of CeCoIn$_{5}$ B =
76 GPa,\cite{Normile} we estimate that a rigid shift of the lattice
parameters for $x=0.83$ and $x=0.9$ corresponds to 0.03 GPa of applied
pressure, Given that $\Delta $T$_{C}$ = 0.9 K (50\%) between these two
concentrations, this result implies that chemical pressure effects are less
relevant than electronic tuning through increased hybridization.\cite%
{Sidorov} This result applies not only at the magnetic - superconducting
boundary where the contribution is only up to 2\%, but also for the whole
range of $x$ since changes of the ground state in Nd$_{1-x}$Ce$_{x}$CoIn$%
_{5} $ are far more dramatic than what was observed in pressure - induced
changes of the ground state in CeCoIn$_{5}$.\cite{Sidorov}

It would be instructive to compare our results with the reported coexistence
of superconductivity and magnetism in CeRhIn$_{5}$ under high pressure and
to field-induced magnetic order in the superconducting state of La$_{1.9}$Sr$%
_{0.1}$CuO$_{4}$.\cite{Tuson}$^{,}$\cite{Knebel}$^{,}$\cite{Lake} In both
materials, as well as in CeCoIn$_{5}$, there are nodes in the
superconducting gap. Delicate balance between antiferromagnetic and
superconducting coupling near a quantum phase transition is smoothly tuned
by magnetic field which generates field-induced vortices that suppress
superconductivity and enhance magnetic correlations.\cite{Demler}$^{,}$\cite%
{Zhang} In addition to this, in Nd$_{1-x}$Ce$_{x}$CoIn$_{5}$ suppression of
d-wave superconductivity is certainly influenced by disorder, as seen in
large increase in residual resistivity (Fig.5).\cite{Balatsky}

The evolution of superconducting T$_{C}$(x) with the increase of Nd is
similar to Ce$_{1-x}$La$_{x}$CoIn$_{5}$.\cite{Petrovic}$^{,}$\cite{Nakatsuji}
This is consistent with dominant potential scattering and insensitivity of T$%
_{C}$ to magnetic configuration of the rare earth ion, as observed in Ce$%
_{1-x}$R$_{x}$CoIn$_{5}$.\cite{Johnpierre}

\section{Conclusion}

In conclusion, we have reported on Nd substitution in CeCoIn$_{5}$. Nd
substitution results in the rich phase diagram, controlled apparently only
by Ce/Nd ratio, i.e. by the level of 4f - conduction electron coupling.
Ground state of this alloy series evolves from LMM\ to HFSC state via
magnetically ordered heavy fermion ground state. Small concentration of Nd
(5\%) induces magnetic order deep in the superconducting state. We have
demonstrated magnetic field tuning of the delicate balance between
superconducting and magnetic ground state at the magnetic - superconducting
boundary. We invite further investigation into possible microscopic
coexistence of magnetic and superconducting order parameters by NMR and
neutron scattering measurements. Possible existence of short range order
(spin glass) between local moment magnetism\ and heavy fermion
superconductor near the middle of alloy series would offer model electronic
system with rich interplay of superconductivity and magnetism comparable to
high-T$_{C}$ cuprate oxide superconductors.

We thank R. Prozorov, S. L. Bud'ko, P. C. Canfield, Johnpierre Paglione and
Myron Strongin for useful communication. This work was carried out at the
Brookhaven National Laboratory, which is operated for the U.S. Department of
Energy by Brookhaven Science Associates (DE-Ac02-98CH10886). This work was
supported by the Office of Basic Energy Sciences of the U.S. Department of
Energy.

\end{document}